# Micro-reflectance and transmittance spectroscopy: a versatile and powerful tool to characterize 2D materials

*Riccardo Frisenda[1], Yue Niu[1,2], Patricia Gant[1], Aday J. Molina-Mendoza,[1] Robert Schmidt[3], Rudolf Bratschitsch[3], Jinxin Liu,[4] Lei Fu,[4] Dumitru Dumcenco[5,6], Andras Kis[5,6], David Perez De Lara[1] and Andres Castellanos-Gomez[1],\**

[1] *Instituto Madrileño de Estudios Avanzados en Nanociencia (IMDEA Nanociencia), Campus de Cantoblanco, E-28049 Madrid, Spain.*
[2] *National Key Laboratory of Science and Technology on Advanced Composites in Special Environments, Harbin Institute of Technology, Harbin, China*
[3] *Institute of Physics and Center for Nanotechnology, University of Münster, 48149 Münster, Germany*
[4] *Laboratory of Advanced Nanomaterials, College of Chemistry and Molecular Science. Wuhan University. Wuhan. China*
[5] *Electrical Engineering Institute, École Polytechnique Fédérale de Lausanne (EPFL), CH-1015 Lausanne, Switzerland*
[6] *Institute of Materials Science and Engineering, École Polytechnique Fédérale de Lausanne (EPFL), CH-1015 Lausanne, Switzerland*

[andres.castellanos@imdea.org](andres.castellanos@imdea.org)

ABSTRACT

Optical spectroscopy techniques such as differential reflectance and transmittance have proven to be very powerful techniques to study 2D materials. However, a thorough description of the experimental setups needed to carry out these measurements is lacking in the literature. We describe a versatile optical microscope setup to carry out differential reflectance and transmittance spectroscopy in 2D materials with a lateral resolution of ~1 μm in the visible and near-infrared part of the spectrum. We demonstrate the potential of the presented setup to determine the number of layers of 2D materials and to characterize their fundamental optical properties such as excitonic resonances. We illustrate its performance by studying mechanically exfoliated and chemical vapor-deposited transition metal dichalcogenide samples.

The use of optical microscopy based characterization techniques has been strongly bound to the born and growth of the field of 2D materials.[1–5] These techniques are widely used since they are fast and simple to implement and, very importantly, they are non-destructive. Quantitative studies of the optical contrast of flakes deposited on $SiO_2$/Si substrates have been firstly used to provide a coarse estimation of the thickness of mechanically exfoliated flakes.[3,6–8] Raman [2,9] and photoluminescence spectroscopy [4,5,10–13] have proven to be very powerful techniques to determine the number of layers of 2D materials and to study their intrinsic optical properties.

Differential reflectance has been also recently used to study the optical properties of 2D materials, especially of transition metal dichalcogenides (TMDCs).[12,14–17] This technique provides a powerful tool to characterize 2D materials (even those that do not present strong excitonic features or Raman modes) in a broad range of the electromagnetic spectrum. Nevertheless, a thorough description of optical setups to characterize 2D materials by differential reflectance is still lacking in the literature. This is precisely the aim of this work: to accurately describe a versatile optical microscopy setup developed to carry out differential reflectance and transmittance measurements on 2D materials with ~1 μm spatial resolution in the wavelength range of 400 nm to 900 nm (1.4 eV to 3.1 eV). The entire system can be easily replicated with a relatively low investment (<9000 €). Moreover, existing optical microscopy setups could be easily upgraded to carry out these spectroscopic measurements.

**Components of the experimental setup:**

Figure 1a shows a photograph of the micro-reflectance/transmittance setup, which we have developed to characterize 2D materials. The setup consists of a Motic BA310 metallurgical microscope, a modified trinocular port, a halogen light source with a liquid light guide and a fiber-coupled CCD spectrometer. Figure 1b shows a detailed photograph of the modified trinocular, indicating the different components employed for its assembly. A 90:10 beam splitter is connected to the C-mount trinocular in order to divide the trinocular light beam into two paths. In one on the paths (10% of the intensity) a CMOS camera is placed at the image plane of the optical system to acquire images of the studied sample. In the other path (90% of the intensity) we place a multimode fiber, also at the image plane, projecting an image of the studied sample on the surface of the fiber end. Table 1 summarizes the complete list of components necessary to assemble the experimental setup, including their part number, distributor, and purchase price.

**Setting up the microscope:**

The size of the fiber core acts as a confocal pinhole, thus the fiber only collects the light coming from a small region of the sample. In order to determine the area and position of that region one can connect the free end of the multimode fiber to a white light source (e.g. the halogen light source) yielding an image of the fiber end facet in the sample plane. The image of the fiber core at the end facet is projected according to the magnification of the microscope lenses and objective assembly. Figure 1c shows the relationship between the core diameter of different multimode fibers and their apparent size in the sample plane. An XY stage allows one to align the image of the end facet with the digital crosshair of the camera that marks the center of the sample image. With reversed beam path (see Figure 1c), this corresponds to the light collection spot and it is possible to determine where the spectra are collected once the light source is disconnected from the fiber end. A focusing stage in the trinocular also allows one to accurately place the fiber core end right in the image plane by moving the focusing stage until the projected fiber core image and the surface of the studied sample are both perfectly in focus. Once the size for light collection has been centered in the crosshair and focused, the fiber is disconnected from the white light source and the free end of the fiber is now connected to the fiber-coupled CCD spectrometer.

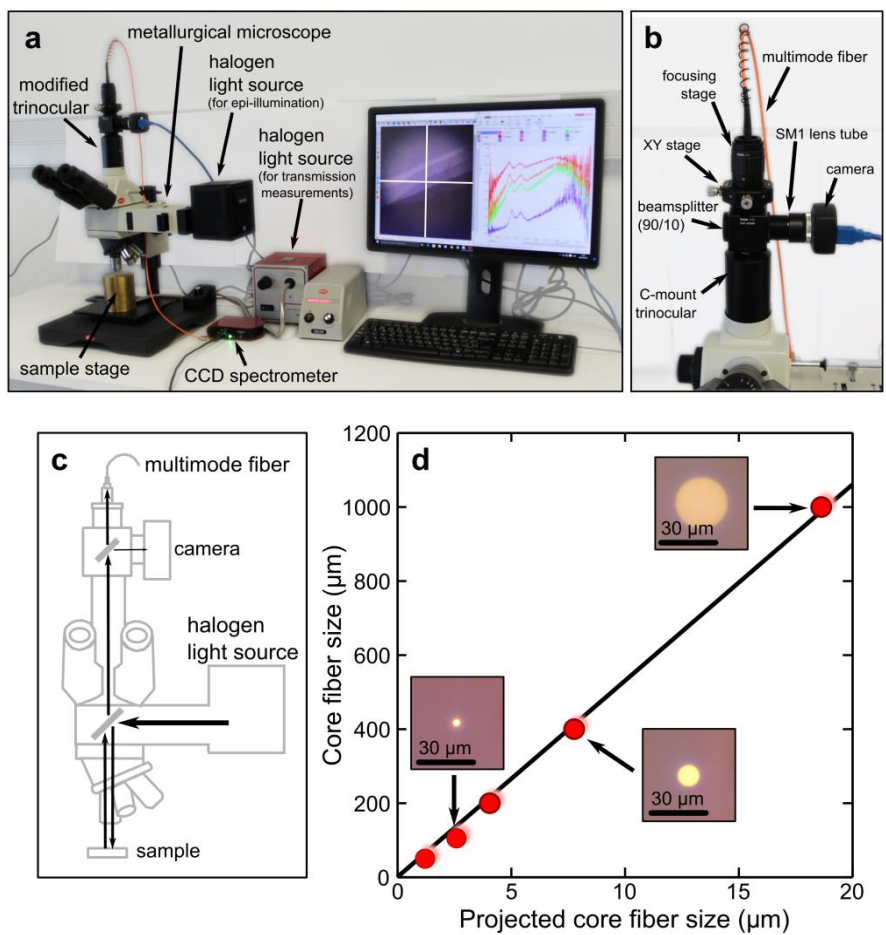

**Figure 1.** (a) Photograph of the experimental setup to perform micro-reflectance and transmittance spectroscopy of 2D materials, highlighting some of the key components. (b) Zoomed photograph of the modified trinocular. (c) Schematic diagram of the optical path of the setup operated in reflection mode. (d) Relationship between the size of the fiber core and its image on the sample surface due to the magnification of the optical system.

**Acquisition of micro-reflectance or transmittance spectra:**

To acquire a micro-reflectance spectrum the microscope is operated in epi-illumination mode. The white light of the microscope halogen lamp is reflected on the sample and collected through a long working distance (infinity corrected) 50× microscope objective (NA 0.55) and directed to the modified trinocular. In epi-illumination operation, closing the field aperture of the microscope (to ~60 µm in diameter) reduces the stray light collected by the finite numerical aperture of the objective lens. Transmittance spectra can be also acquired by turning off the microscope lamp and placing the sample onto a sample holder connected to the halogen light source by a fiber bundle (NA 0.57). The sample holder has a see-through hole where the fiber bundle can be slid in. The sample is directly placed onto the end facet of the fiber bundle without using other lenses (see Supporting Information).

| Part number | Description | Distributor | Price (€) |
|---|---|---|---|
| BA310Met-H Trinocular | Metallurgical microscope (with epi-illumination) | Motic | 3620.00 |
| CCS200/M | CCD spectrometer | Thorlabs | 2475.00 |
| OSL2 | Halogen lamp with fiber bundle | Thorlabs | 807.30 |
| SM1SMA | Adaptor to connect the multimode fiber to the halogen lamp | Thorlabs | 25.22 |
| Adaptor BA Series | C-mount threaded trinocular | Motic | 127.00 |
| SM1A10 | Adaptor to connect the beam splitter mount to the C-mount trinocular | Thorlabs | 16.88 |
| CCM1-4ER/M | Mount for beam splitter cube | Thorlabs | 115.20 |
| BS025 | 90:10 beam splitter cube | Thorlabs | 186.30 |
| SM1A10 | Adaptor to connect the beam splitter mount to the XY stage | Thorlabs | 16.88 |
| CXY1 | XY stage | Thorlabs | 152.15 |
| SM1NR05 | Focusing stage | Thorlabs | 162.00 |
| SM05SMA | Adaptor to connect the multimode fiber to the focusing stage | Thorlabs | 24.57 |
| M96L01 | 105 µm core multimode fiber | Thorlabs | 95.53 |
| SM1V10 | SM1 lens tube to connect to the beam splitter mount | Thorlabs | 29.34 |
| SM1A39 | Adaptor to connect the SM1 lens tube to the CMOS camera | Thorlabs | 18.00 |
| #89-734 | CMOS camera | Edmund Optics | 563.50 |
| --- | Homemade brass sample holder with hole for the fiber bundle | Workshop | 150.00 |
| | | **Total** | **8584.87** |

**Table 1:** List of the components employed to assemble the experimental setup.

**Experimental results on monolayer and few-layer TMDCs:**

Figure 2a shows a reflection mode optical image of a MoS$_2$ flake deposited on a PDMS substrate (gel-film from Gel-Pak®) by mechanical exfoliation of bulk MoS$_2$ (Moly Hill mine, Quebec, QC, Canada) with Nitto tape (Nitto Denko Co., SPV224 clear). The single-layer region of the flake has been highlighted with a dashed line to facilitate its identification. Figure 2b shows the spectra acquired at the positions marked in the optical image of Figure 2a with a blue square (flake, $R$) and a red circle (substrate, $R_0$). The differential reflectance spectrum is then calculated as ($R$-$R_0$)/$R$ (see Figure 2c) and it is related to the absorption coefficient of the material α(λ) as[14,18]

$$\frac{R - R_0}{R} = \frac{4n}{n_0^2 - 1} \alpha(\lambda)$$

where $n$ is the refractive index of the flake under study and $n_0$ is the refractive index of the substrate. Figure 2d shows a transmission mode optical image of the same flake shown in Figure 2a. Figure 2e shows the transmission spectra acquired at the positions highlighted with a blue square (flake, $T$) and red circle (substrate, $T_0$) in Figure 2d. The flake transmittance can be directly determined by dividing $T/T_0$ (see Figure 2f).

Both the differential reflectance and transmittance spectra have two prominent narrow peaks/dips occurring at wavelengths ~1.90 eV and ~2.05 eV that correspond to the optical absorption due to the direct transitions at the K point of the Brillouin zone, associated to the generation of the A and B excitons, respectively. This observation is in agreement with previous photoluminescence studies.[4,5,19] The inset in Figure 2f shows a simplified band diagram at the K point of the Brillouin zone, where the origin of the A and B excitonic resonances is illustrated. The valence band is split because of the strong spin-orbit interaction due to the Mo atoms giving rise to two direct band gap transitions at the K point. The spectra also show a broad feature around 2.85 eV. This feature is typically not observed in photoluminescence experiments, which mostly use an excitation wavelength of ~2.3 eV. Recent hyperspectral, reflectance, ellipsometry and photocurrent spectroscopy experiments, however, present this feature (referred to as the C exciton peak) It originates from singularities in the joint density of states between the first valence and conduction bands near the Γ point of the valence band that leads to multiple optical transitions nearly degenerate in energy.[14,15,20–24]

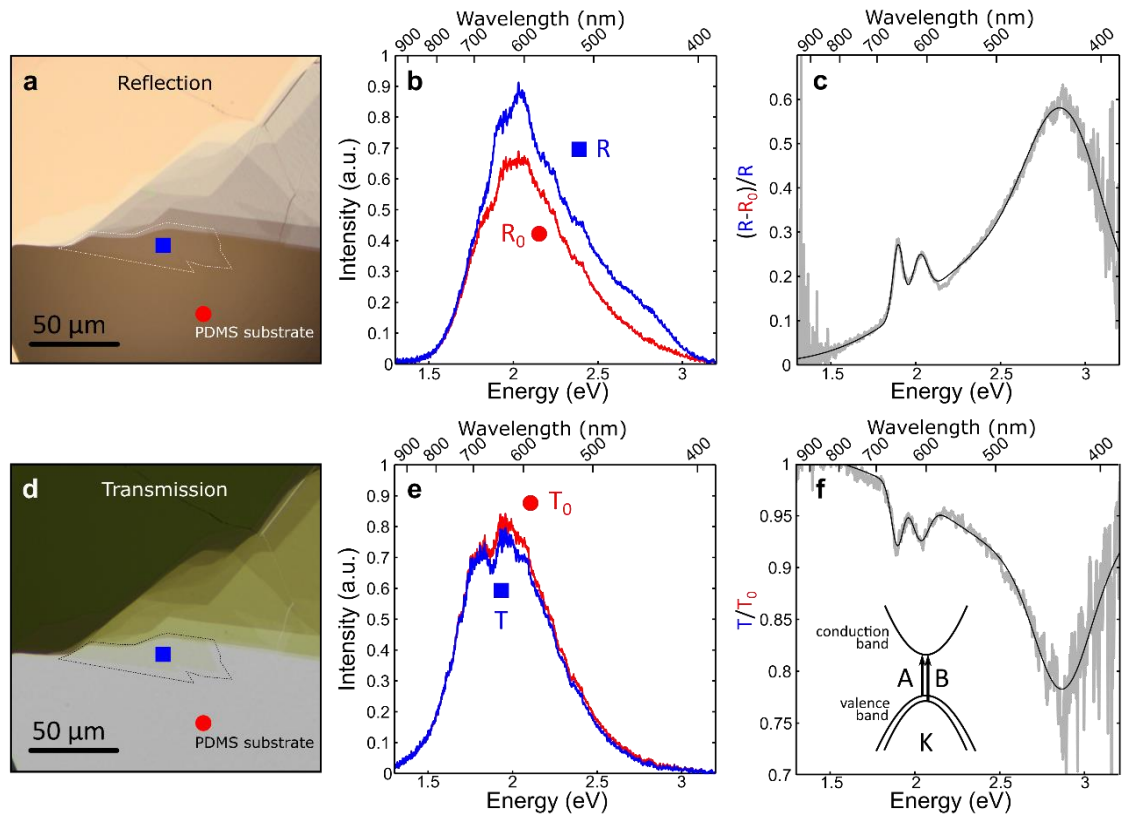

**Figure 2.** (a) Optical reflection image of a $MoS_2$ flake transferred onto a PDMS substrate (epi-illumination). The single layer area has been highlighted with a dashed line. (b) Reflection spectra acquired at the positions marked with the blue square and red circle in (a). (c) Calculated differential reflectance of the single layer $MoS_2$ from the spectra displayed in (b). The thin black line is a multi-Gaussian fit used to extract the peak position and full width half maximum values of the different excitons. (d) Transmission mode optical image of the same flake shown in (a). (e) Transmission spectra acquired on the flake (blue square) and substrate (red circle), employed to calculate the transmittance spectrum of the single layer flake (f). The inset in (f) shows a simplified schematic of the band diagram around the K point, illustrating the origin of the two excitonic resonances at 1.90 eV and 2.05 eV.

Micro-reflectance and transmittance are interesting techniques to determine the number of layers of 2D materials. Figure 3a shows, as an example, the differential reflectance spectra acquired for $MoS_2$ flakes with thickness ranging from 1 to 7 layers. One can determine the number of layers from the energy of the A exciton, which blue-shifts upon reduction of the number of layers (see Figure 3b).[25] The B exciton, on the other hand, shows a weaker thickness dependence.[25] These results are in good agreement with previous photoluminescence measurements.[5] Interestingly, the C exciton peak also shifts with the number of layers and its dependence is even stronger than that of the A exciton, thus one can use the position of the C exciton peak to determine independently the number of layers of $MoS_2$.[14] Note that photoluminescence measurements are typically limited to study the A and B excitons, as the excitation laser employed is typically in the 2.0-2.3 eV range.[4,16] Figure 3b summarizes the thickness dependence of the A, B, and C exciton resonances. A more comprehensive study of the thickness dependence excitonic features in W- and Mo-based TMDCs will be published elsewhere.

The micro-reflectance measurements of $MoS_2$ flakes with of different thickness (Figure 3a to 3c) are compared with results obtained with Raman spectroscopy. Raman spectroscopy is a well-established technique to determine the number of layers of $MoS_2$, as the frequency difference between the two most prominent Raman modes (the $A_{1g}$ and the $E^1_{2g}$ peaks) increases monotonically with the number of layers.[9,26] Figure 3d shows the Raman spectra acquired on the $MoS_2$ flakes on PDMS, with different number of layers. The thickness dependence of the two modes is displayed in Figure 3e. Figure 3f depicts the frequency difference between them. A direct comparison between Figure 3e and Figures 3b and 3c indicates that micro-reflectance measurements are very effective to determine the number of layers, especially for multilayer flakes (thicker than 4 layers) where the frequency difference between the Raman modes starts to saturate.

It is important to note that micro-reflectance (or transmittance) measurements are very powerful to investigate indirect band gap semiconductors that typically require long exposure times in order to be studied by photoluminescence. In fact, the quantum yield of $MoS_2$ strongly decreases by more than a factor of 1000 from the single layer case (direct band gap) to the multilayer case (indirect band gap).[5] Moreover, these techniques can be used to study the optical properties of conducting and insulating materials without strong excitonic features, which constitutes another important advantage with respect to photoluminescence.

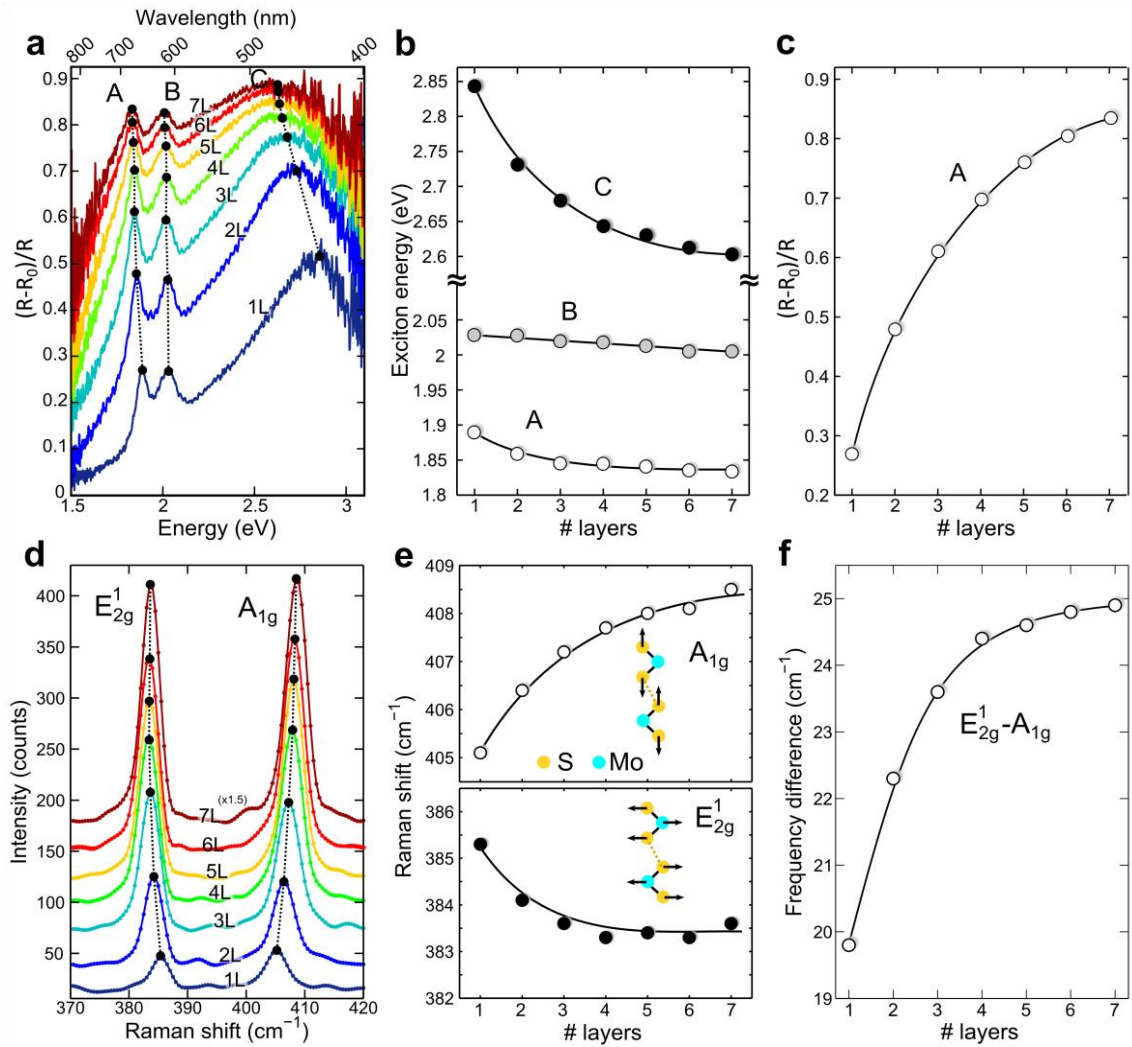

**Figure 3.** (a) Differential reflectance spectra acquired for $MoS_2$ flakes with different number of layers. The peak positions of the different excitons are highlighted by filled circles. (b) Thickness dependence of the excitons energy. (c) Amplitude of the A exciton peak as a function of the number of layers. (d) Raman spectra acquired for $MoS_2$ flakes with different number of layers. The peak positions of the two most prominent Raman modes are highlighted by filled circles. (e) Thickness dependence of the Raman shift of the $A_{1g}$ and $E^1_{2g}$ modes. (f) Frequency difference between the Raman modes as a function of the number of layers.

In order to illustrate the flexibility and versatility of this technique to identify and to study other 2D materials we have acquired differential reflectance spectra for single-layer flakes of other transition metal dichalcogenides. Figure 4 shows a comparison between the differential reflectance spectra of single-layer $WS_2$, $MoS_2$, $WSe_2$ and $MoSe_2$. The excitonic resonances have been labelled according to the standard notation employed in the literature. One can easily identify the material from the position of the A and B exciton peaks (see Table 2), which are in very good agreement with the reported values from photoluminescence experiments.[4,17,27,28] The origin of these A and B excitonic resonance peaks in $WS_2$, $WSe_2$ and $MoSe_2$ is very similar to the one in $MoS_2$, i.e. the direct band gap transitions at the K point of the Brillouin zone (as sketched in the inset of Figure 2f). Therefore, one can estimate the magnitude of the spin-orbit interaction in these materials from the energy difference between the A and B exciton. For W-based TMDCs this value is larger due to the higher atomic number of W with respect to Mo. Similarly, for the same transition metal, the Se- based TMDCs have larger spin-orbit interaction than S-based ones. The C peak observed in $WS_2$ and $MoSe_2$ is believed to have the same origin as the one observed in $MoS_2$.[14,15,20–24,29] In case of $WSe_2$ several transitions along the $\overline{K\Gamma}$ direction contribute to the C exciton. The D exciton transitions stem from the same bands as the B exciton, but occur along the $\overline{KM}$ direction [17,28].

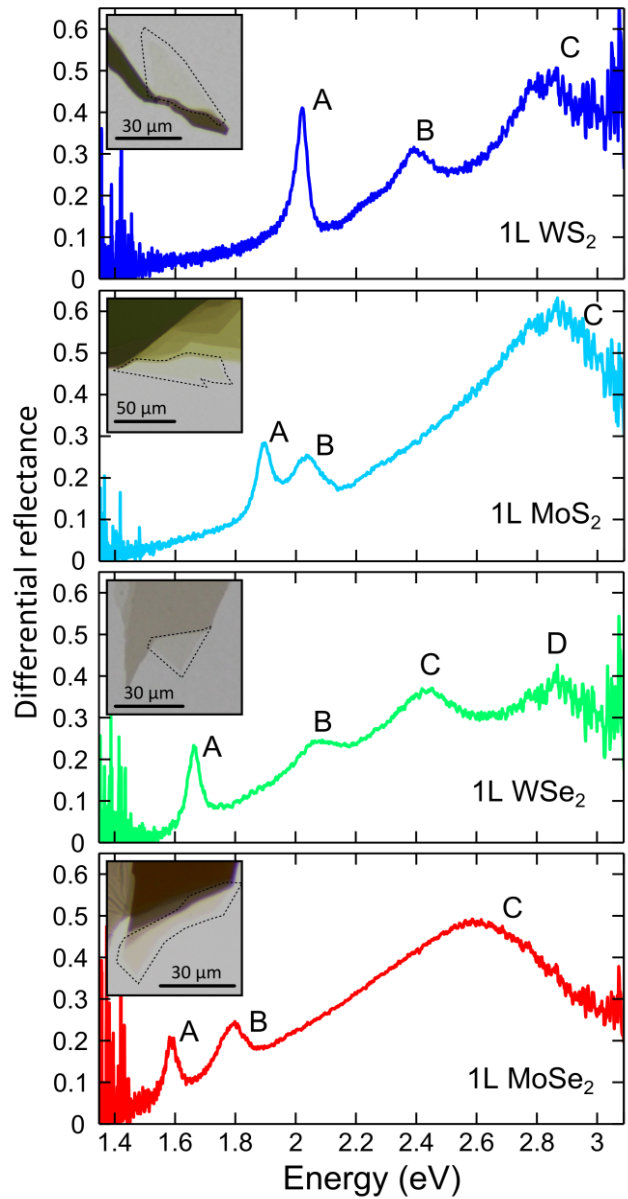

**Figure 4.** Differential reflectance spectra acquired for single-layer flakes of different transition metal dichalcogenides: $WS_2$, $MoS_2$, $WSe_2$ and $MoSe_2$ (from top to bottom). The exciton resonances have been highlighted according to the notation employed in the literature.[15,28] (Insets) Transmission mode optical images of the studied single-layer flakes (highlighted with a dashed line).

| Material | A exciton (eV) | B exciton (eV) |
|---|---|---|
| 1L $WS_2$ | 2.02 | 2.39 |
| 1L $MoS_2$ | 1.90 | 2.04 |
| 1L $WSe_2$ | 1.66 | 2.06 |
| 1L $MoSe_2$ | 1.59 | 1.80 |

**Table 2:** Summary of the A and B exciton energies extracted from the differential reflectance measurements shown in Figure 4.

Finally, we demonstrate the suitability of differential reflectance to characterize 2D materials prepared by large area growth methods such as chemical vapor deposition (CVD). Figure 5a compares the differential reflectance spectra acquired for monolayers of $MoS_2$ prepared by mechanical exfoliation and by CVD growth. Two different sets of CVD-prepared samples have been characterized (see the Supporting Information for details about the growth methods). One has been grown on a highly polished sapphire substrate resulting in characteristic single-crystal domains in the shape of well-defined equilateral triangles.[30] The other CVD sample has been directly grown on a quartz substrate. The energy of the excitonic peaks of the CVD samples is consistent with the ones measured on mechanically exfoliated single-layer $MoS_2$, although slightly red-shifted. This red-shift in CVD-grown samples is typically attributed to strain induced during the thermal cycling during growth.[31] Differential reflectance spectroscopy is therefore a powerful tool to study also CVD-grown samples. Moreover, the fast measurement time (approximately 1 second per spectrum) allows one to statistically analyze the quality of a CVD-grown sample by acquiring spectra over a large number of sample positions. Figure 5b and 5c show histograms of the energy and the full width half maximum of the A exciton peak measured on 116 different $MoS_2$ triangular shaped crystallites grown on the sapphire substrate, demonstrating the high uniformity of the CVD sample.

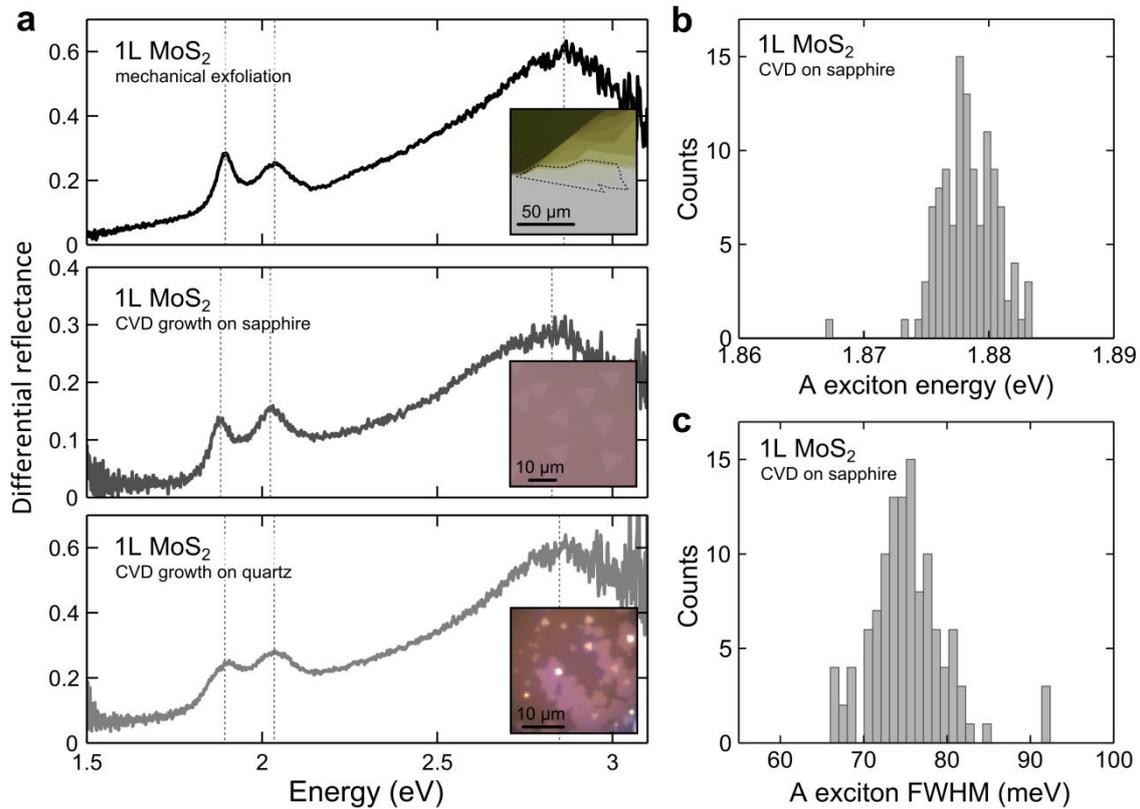

**Figure 5.** Differential reflectance spectra acquired on single-layer $MoS_2$ samples prepared by mechanical exfoliation on PDMS (top panel in (a)), CVD growth on sapphire (middle panel in (a)) and CVD growth on quartz (bottom panel in (a)). The insets in (a) show optical images of the studied samples. (b) and (c) show the statistical analysis of the energy and full-width at half maximum of the A exciton peak measured over 116 $MoS_2$ crystallites on the CVD sample grown on sapphire, showing a remarkable uniformity of the CVD grown material.

## Conclusions:

In summary, we have presented a simple and versatile optical microscope setup to carry out micro-reflectance and transmittance spectroscopy with a lateral resolution of ~1 μm in the spectral range of 400 nm to 900 nm. The setup can be used to unambiguously identify different 2D materials and to determine their thicknesses, as well as to characterize their fundamental optical properties. We believe that the presented setup can be easily replicated by other groups, as it is mainly based on standard commercially available optical components that can be purchased for a price of less than 9000 €. Moreover, already existing optical microscope setups can be easily upgraded by purchasing only part of the components.

**Acknowledgements:**

We acknowledge funding from the European Commission under the Graphene Flagship, contract CNECTICT-604391. AC-G acknowledges financial support from the MINECO (Ramón y Cajal 2014 program RYC-2014- 01406 and MAT2014-58399-JIN) and from the Comunidad de Madrid (MAD2D-CM Program (S2013/MIT-3007)). RF acknowledges support from the Netherlands Organisation for Scientific Research (NWO) through the research program Rubicon with project number 680-50-1515. DPdL acknowledges support from the MINECO (program FIS2015-67367-C2-1-P). YN acknowledges the grant from the China Scholarship Council (File NO. 201506120102).

# Supporting Information

## Micro-reflectance and transmittance spectroscopy: a versatile and powerful tool to characterize 2D materials

*Riccardo Frisenda[1], Yue Niu[1,2], Patricia Gant[1], Aday J. Molina-Mendoza,[1] Robert Schmidt[3], Rudolf Bratschitsch[3], Jinxin Liu,[4] Lei Fu,[4] Dumitru Dumcenco[5,6], Andras Kis[5,6], David Perez De Lara[1] and Andres Castellanos-Gomez[1],\**

[1] *Instituto Madrileño de Estudios Avanzados en Nanociencia (IMDEA Nanociencia), Campus de Cantoblanco, E-28049 Madrid, Spain.*
[2] *National Key Laboratory of Science and Technology on Advanced Composites in Special Environments, Harbin Institute of Technology, Harbin, China*
[3] *Institute of Physics and Center for Nanotechnology, University of Münster, 48149 Münster, Germany*
[4] *Laboratory of Advanced Nanomaterials, College of Chemistry and Molecular Science. Wuhan University. Wuhan. China*
[5] *Electrical Engineering Institute, École Polytechnique Fédérale de Lausanne (EPFL), CH-1015 Lausanne, Switzerland*
[6] *Institute of Materials Science and Engineering, École Polytechnique Fédérale de Lausanne (EPFL), CH-1015 Lausanne, Switzerland*

*andres.castellanos@imdea.org*

**Calibration of the optical microscope:**

In order to calibrate the optical microscope setup we acquired optical images of a standard sample. We employed a compact disc (CD) as the distance between the tracks is a well-known distance (1.6 μm).

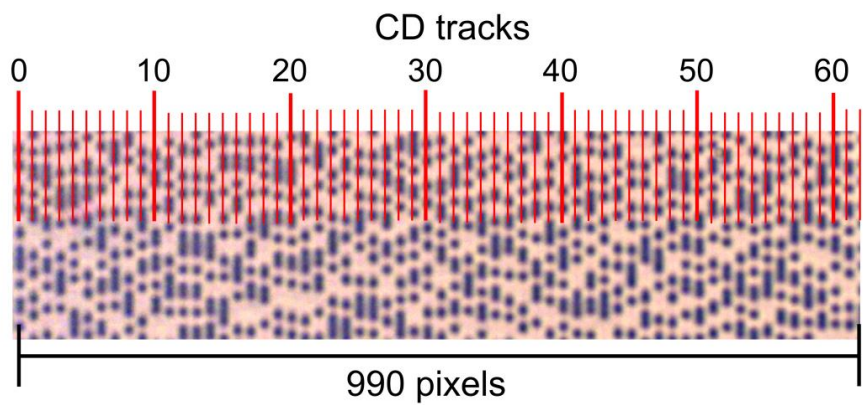

**Figure S1.** Portion of the optical image of a CD simple in epi-illumination mode, employed to calibrate the optical microscope setup.

**Comparison between the transmittance spectra obtained with the presented setup and the spectra obtained by multispectral and hyperspectral imaging:**

Figure S2 shows a comparison between the transmittance spectra of single-layer $MoS_2$ flakes obtained with different experimental setups: a multispectral setup and hyperspectral setup. For the multispectral setup, a conventional optical microscope equipped with a monochrome CMOS camera is employed and the illumination wavelength is selected by simply changing narrow bandwidth bandpass filters. The hyperspectral setup employed a white light source connected to a monochromator to select the illumination wavelength (see more details in Ref. [22]).

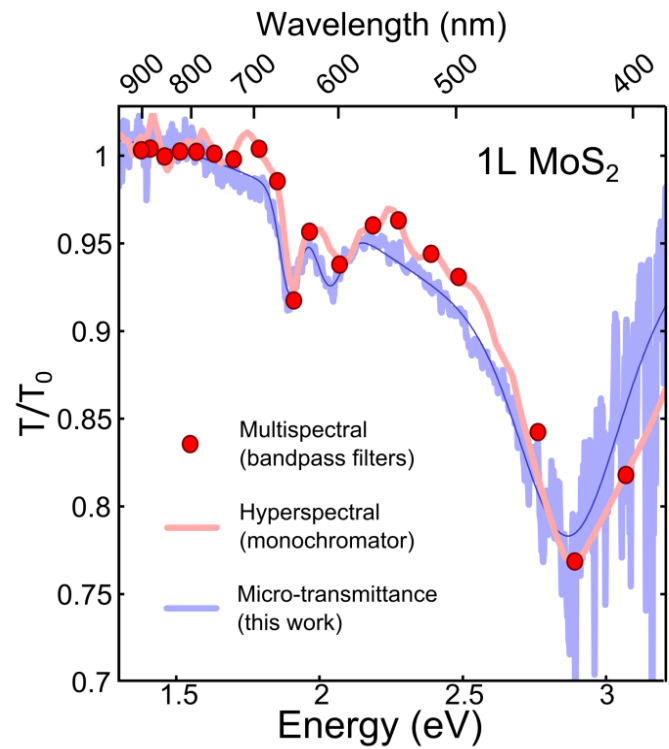

**Figure S2.** Comparison between the transmittance spectra of single-layer $MoS_2$ obtained with the presented setup and the spectra obtained by multispectral and hyperspectral imaging.

**Zoomed in picture of the homemade sample stage with a hole to fit in the fiber bundle connected to the halogen light source:**

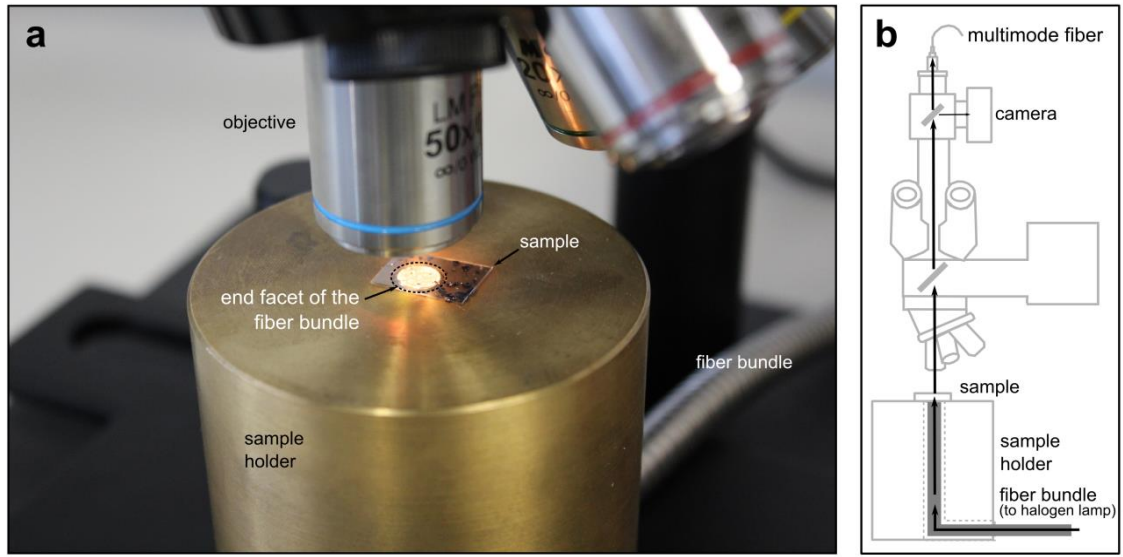

**Figure S3.** Zoomed in picture of the homemade sample stage with a hole to fit in the fiber bundle connected to the halogen light source. The fiber bundle numerical aperture is 0.57.

**Spectra of the white light sources used for reflectance and transmittance measurements:**

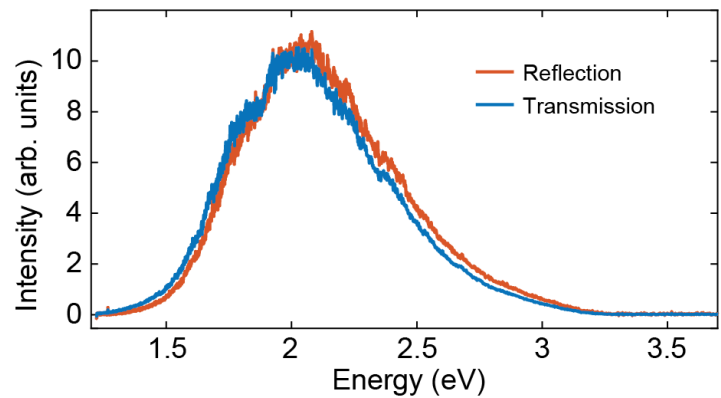

**Figure S4.** The wavelength range of the experimental setup has been determined by directly acquiring a spectrum of the two white illumination halogen sources that we use for reflectance and for transmission measurement.

## CVD growth procedure:

*CVD MoS$_2$ on sapphire:* The monolayer molybdenum disulfide MoS$_2$ samples have been obtained by the chemical vapor deposition (CVD) method on highly polished sapphire substrates. The growth process is based on the gas-phase reaction between MoO$_3$ (≥ 99.998% Alfa Aesar) and high-pure sulfur evaporated from the solid phase (≥ 99.99% purity, Sigma Aldrich). The growth procedure results in characteristic single-crystal domains in the shape of well-defined equilateral triangles that merge into a continuous monolayer film in the middle part of the growth substrate.[S1]

*CVD MoS$_2$ on quartz:* The growth of MoS$_2$ on quartz was carried out in a CVD furnace with 1 inch quartz tube. Before the growth, the quartz substrate was placed at the center of the furnace and the MoO$_3$ powder (Alfa Aesar, 99.9999%, 15 mg) were placed at the upper stream side of the quartz. The S powders (Alfa Aesar, 99.95%, 30 mg) were placed in a quartz boat located at the upper stream . The furnace was first heated up to 650 ℃ at a rate of 20 ℃ per minute with 150 sccm Ar. When the temperature of furnace reached 650 ℃, sulfur was heated to 150 °C. The furnace was then heated up to 850 °C in 15 min and kept this temperature for 30 min. After growth, the furnace was allowed to cool down to the room temperature in pure Ar atmosphere.

## Supporting Information References